\documentclass[preprint,aps,a4,amsmath,amssymb,floatfix]{revtex4-1}

\usepackage{bm}
\usepackage{graphicx}
\usepackage{amsbsy}
\newcommand{\vlk}{$V_{\rm low-k}$ }
\newcommand{\vlkn}{$V_{\rm low-k}$}
\newcommand{\be}{\begin{equation}}
\newcommand{\ee}{\end{equation}}

\begin{document} 

\title{Quasiparticle interaction in nuclear matter with chiral three-nucleon
forces\footnote{Work 
supported in part by BMBF, GSI and by the DFG cluster of excellence: Origin and 
Structure of the Universe.}}

\author{J.\ W.\ Holt, N.\ Kaiser, and W.\ Weise}
\affiliation{Physik Department, Technische Universit\"{a}t M\"{u}nchen,
    D-85747 Garching, Germany}

\begin{abstract}
We derive the effective interaction between two quasiparticles in symmetric 
nuclear matter resulting from the leading-order chiral three-nucleon force. 
We restrict our study to the $L=0,1$ Landau parameters of the central
quasiparticle interaction computed to first order. 
We find that the three-nucleon force provides substantial repulsion in the 
isotropic spin- and isospin-independent component $F_0$ of the 
interaction. This repulsion acts to stabilize nuclear matter against isoscalar density 
oscillations, a feature which is absent in calculations employing low-momentum 
two-nucleon interactions only. We find a rather large uncertainty for
the nuclear compression modulus ${\cal K}$ due to a sensitive
dependence on the low-energy constant $c_3$. The effective nucleon mass $M^*$ on 
the Fermi surface, as well as the nuclear symmetry energy $\beta$, 
receive only small corrections from the leading-order chiral three-body force. Both
the anomalous orbital $g$-factor $\delta g_l$ and the Landau-Migdal parameter 
$g_{NN}^\prime$ (characterizing the spin-isospin response of nuclear matter)
decrease with the addition of three-nucleon correlations. In fact, $\delta g_l$ 
remains significantly smaller than its value extracted from experimental data,
whereas $g_{NN}^\prime$ still compares well with empirical values.
The inclusion of the three-nucleon force results in relatively small $p$-wave 
($L=1$) components of the central quasiparticle interaction, thus suggesting an 
effective interaction of short range.
\end{abstract}

\maketitle

\section{Introduction}

In a recent publication \cite{holt11} we have studied the quasiparticle
interaction in isospin-symmetric nuclear matter employing realistic chiral
nucleon-nucleon (NN) interactions. The second-order calculation performed in 
ref.\ \cite{holt11} explored the order-by-order convergence of the perturbative 
expansion for the quasiparticle interaction as well as the scale dependence 
resulting from the renormalization group evolution \cite{bogner03,bogner10} 
of the underlying two-body interaction. Although good agreement was found 
for a number of nuclear observables, such as the effective mass $M^*$ at the Fermi 
surface, the nuclear symmetry energy $\beta$ and the spin-isospin 
response encoded in the parameter $g_{NN}^\prime$, both the compression
modulus ${\cal K}$ of isospin-symmetric nuclear matter and the anomalous orbital 
$g$-factor $\delta g_l$ were found to differ appreciably from empirical values
extrapolated from collective excitations of finite nuclei. Most seriously, the compression 
modulus ${\cal K}$ of nuclear matter (encoded in the Fermi liquid parameter $F_0$) was 
found to be negative at both first- and second-order across a wide range of 
cutoff scales.

Previous studies \cite{sjoberg73a,sjoberg73b,dickhoff83,backman85,holt07} have
suggested that interactions induced by the polarization of the medium 
(first studied by Babu and Brown for the case of liquid $^3$He \cite{babu73})
provide sufficient repulsion to stabilize nuclear matter against isoscalar density 
fluctuations. In fact, already the leading-order contribution to the Babu-Brown
induced interaction is 
quite repulsive \cite{holt11}, though not enough to achieve stability at normal
nuclear matter saturation density when employing chiral or low-momentum 
NN interactions. For such potentials, numerous calculations 
\cite{fritsch05,bogner05,siu09,hebeler11} of the equation of state
have revealed the necessity of three-nucleon forces in driving saturation 
toward the empirical density $\rho_0 =0.16$\,fm$^{-3}$ and energy per particle $\bar E = 
-16$\, MeV.

To improve the microscopic description of the quasiparticle interaction, we compute in 
the present work the first-order (perturbative) contribution to the 
$L=0,1$ Landau parameters resulting from the 
leading-order chiral three-nucleon force. Previous work \cite{kaiser06}
performed within the framework of chiral effective field theory
included explicit $\Delta$-isobar degrees of freedom in a calculation of the 
isotropic central Fermi liquid parameters. In the present study we employ the 
high-precision Idaho N$^3$LO chiral NN interaction \cite{entem03} together with
the N$^2$LO chiral three-nucleon force \cite{epelbaum06}, which subsumes 
certain processes involving virtual $\Delta$-isobar excitations 
considered in ref.\ \cite{kaiser06}. The additional repulsion provided 
by the leading-order three-nucleon force is expected to play an important role 
for the Landau parameter $F_0$, but the extent to which other 
nuclear observables respond to additional three-nucleon correlations has not 
been studied systematically in calculations employing high-precision NN 
potentials.

The present paper is organized as follows. In Section \ref{qp3n} we briefly 
review Landau's theory of normal Fermi liquids and discuss the connection
between physical observables and various Fermi liquid parameters. We then 
derive analytical expressions for the first-order contribution to the $L=0,1$ 
central Landau parameters arising from the leading-order chiral 
three-nucleon interaction. In section \ref{results} we present numerical 
results for the density and scale dependence of the quasiparticle interaction 
supplemented by the second-order contributions of ref.\ \cite{holt11}. 
The latter were derived from the Idaho N$^3$LO chiral 
two-nucleon interaction as well as from the (universal)
low-momentum NN interaction \vlkn. We end with a summary and an outlook.


\section{Nuclear quasiparticle interaction}
\label{qp3n}

\subsection{Landau parameters and nuclear observables}

Fermi liquid theory was introduced by Landau in the 1950's \cite{landau57} to 
describe the properties of strongly interacting {\it normal} many-fermion
systems at low 
temperatures. The low-energy excitations about the ground 
state are long-lived quasiparticles that retain certain 
features of non-interacting (independent) particles but have
modified dynamical properties such an effective mass and an effective
magnetic moment. Fermi liquid theory has been used to describe a wide
variety of quantum many-body systems, including various types of conductors at low 
temperatures, liquid $^3$He, nuclear matter, neutron matter and also
finite nuclei \cite{migdal67,baym91,nozieres99}.

The quasiparticle interaction ${\cal F}(\vec p_1 \sigma_1 \tau_1; \vec p_2 \sigma_2 \tau_2)$ 
encodes bulk equilibrium and transport properties
of a Fermi liquid. It arises at second order in the expansion of the energy
density in terms of powers of the quasiparticle distribution function 
$\delta n_{\vec p \sigma \tau}$:
\be
\delta {\cal E} = \sum_{\vec p, \sigma, \tau} \epsilon_{\vec p}\, \delta 
n_{\vec p \sigma \tau} + \frac{1}{2} \sum_{\substack{{\vec p}_1\sigma_1\tau_1
 \\ {\vec p}_2\sigma_2\tau_2}}{\cal F}({\vec p}_1 \sigma_1\tau_1;
{\vec p}_2\sigma_2\tau_2) \delta n_{{\vec p}_1 \sigma_1 \tau_1}
\delta n_{{\vec p}_2 \sigma_2 \tau_2} + \cdots,
\label{deltae}
\ee
where $\epsilon_{\vec p}$ is the single-particle energy and higher-order
quasiparticle correlations are neglected.
The central part of the quasiparticle interaction in spin- and isospin-saturated 
nuclear matter has the form
\begin{equation}
{\cal F}({\vec p}_1, {\vec p}_2) =
f({\vec p}_1,{\vec p}_2) + f^\prime({\vec p}_1,{\vec p}_2) {\vec \tau}_1 \cdot
{\vec \tau}_2 + \left [g({\vec p}_1,{\vec p}_2) + g^\prime({\vec p}_1,
{\vec p}_2) {\vec \tau}_1 \cdot {\vec \tau}_2\right ] {\vec \sigma}_1 \cdot 
{\vec \sigma}_2 \, ,
\label{ffunction}
\end{equation}
where $\vec \sigma_{1,2}$ and $\vec \tau_{1,2}$ denote the spin and 
isospin operators of the two nucleons on the Fermi surface $|\vec p_1\,| = |\vec p_2\,| = k_F$. 
More generally, the
quasiparticle interaction can include (non-central) tensor components which modify the 
stability conditions of nuclear matter \cite{backman79} and are necessary for 
understanding the magnetic susceptibility \cite{haensel82} and
the response of nuclear matter to weak probes 
\cite{olsson04,bacca09}. For two quasiparticles on the Fermi surface, the remaining
angular dependence of their interaction can be expanded in Legendre polynomials
of $\cos \theta ={\hat p}_1 \cdot {\hat p}_2$:
\begin{equation}
X({\vec p}_1,{\vec p}_2) = \sum_{L=0}^\infty X_L(k_F) P_L(\mbox{cos } \theta),
\label{gflp}
\end{equation}
where $X$ represents $f, f^\prime, g,$ or $g^\prime$, and the angle $\theta$ is
related to the relative momentum $p = \frac{1}{2} |{\vec p}_1 - {\vec p}_2|$ 
through the relation $p = k_F\, {\rm sin}\, (\theta /2)$.
The coefficients of the expansion in eq.(\ref{gflp}) are referred to as the 
Fermi liquid parameters (FLPs). It is conventional to introduce dimensionless 
FLPs by multiplying by the density
of states at the Fermi surface, $N_0=2M^*k_F/\pi^2$,
where $M^*$ is the nucleon effective mass and $k_F$ is the Fermi momentum, 
leading to
\begin{equation}
{\cal F}({\vec p}_1, {\vec p}_2)=\frac{1}{N_0}\sum_{L=0}^\infty \left [
F_L + F^\prime_L \vec \tau_1 \cdot
\vec \tau_2 + (G_L + G^\prime_L \vec \tau_1 \cdot \vec \tau_2) \vec \sigma_1 
\cdot \vec \sigma_2 \right ]P_L({\rm cos}\, \theta).
\label{ffunction2}
\end{equation}
For short-range interactions the expansion in (\ref{ffunction2}) is typically rapidly
converging such that only a few constants characterize the dynamics of low-energy
excitations. Moreover, individual parameters evaluated at the equilibrium Fermi 
momentum $k_F=1.33$\,fm$^{-1}$ are related to properties 
of the quasiparticles and the bulk nuclear medium:
\begin{eqnarray}
{\rm Quasiparticle \, \, effective \, \, mass:} \hspace{.2in} \frac{M^*}{M_N} &=& 1+F_1/3, 
\nonumber \\
{\rm In\mbox{-}medium \, \, orbital \, \,} g\mbox{-factor:} \hspace{.2in} 
g_l &=& \frac{1+\tau_3}{2} + \frac{F^\prime_1-F_1}{6(1+F_1/3)}\tau_3 , \nonumber \\
{\rm Compression \, \, modulus:} \hspace{.2in} {\cal K}&=&\frac{3k_F^2}{M^*} 
\left (1+F_0\right ), \nonumber \\
{\rm Isospin\, \,  asymmetry \, \, energy:} \hspace{.2in} \beta &=& 
\frac{k_F^2}{6M^*}(1+F_0^\prime), \nonumber \\
{\rm Spin\mbox{-}isospin\, \, response:} \hspace{.2in} g_{NN}^\prime &=& 
\frac{4M_N^2}{g_{\pi N}^2N_0} G_0^\prime,
\label{obs}
\end{eqnarray}
where the anomalous orbital $g$-factor $\delta g_l$ is given by
\be
\delta g_l = \frac{F^\prime_1-F_1}{6(1+F_1/3)}
\ee
and $g_{\pi N} \simeq 13.2$ is the strong $\pi N$ coupling constant.
Spin observables, though largely unconstrained experimentally, receive 
significant contributions from the (non-central) tensor Fermi liquid parameters 
\cite{haensel82}.

\begin{figure}
\begin{center}
\includegraphics[height=4cm]{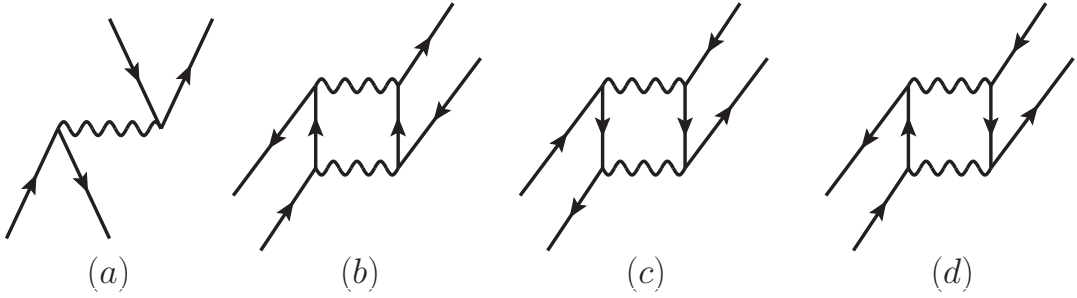}
\end{center}
\vspace{-.5cm}
\caption{First- and second-order diagrammatic contributions to the quasiparticle
interaction. The wavy lines symbolize the two-nucleon interaction.}
\label{qp1st2nd}
\end{figure}

Recently, we have carried out systematic calculations \cite{holt11} of the 
quasiparticle interaction in nuclear matter to second order in many-body perturbation 
theory employing chiral and low-momentum NN interactions. The first- and 
second-order contributions, shown diagrammatically in Fig.\ \ref{qp1st2nd}, 
have the form
\begin{equation}
{\cal F}^{(1)}({\vec p}_1 s_1 t_1; {\vec p}_2 s_2 t_2)
 = \langle {\vec p}_1 s_1 t_1; {\vec p}_2 s_2 t_2 | 
\bar V | {\vec p}_1 s_1 t_1; {\vec p}_2 s_2 t_2 \rangle
\equiv \langle 12 | \bar V | 12 \rangle
\label{order1qp}
\end{equation}
and
\begin{eqnarray}
{\cal F}^{(2)} ({\vec p}_1 s_1 t_1; {\vec p}_2 s_2 t_2)
&=& \frac{1}{2} \sum_{34} \frac{|\langle 12 | \bar V | 34 \rangle|^2 (1-n_3)
(1-n_4)} {\epsilon_1 + \epsilon_2 - \epsilon_3 - \epsilon_4} \nonumber \\
 &+& \frac{1}{2} \sum_{34} \frac{|\langle 12 | \bar V | 34 \rangle |^2 n_3 n_4}
{\epsilon_3 + \epsilon_4 - \epsilon_1 - \epsilon_2} 
 - 2 \sum_{34} \frac{|\langle 13 | \bar V | 24 \rangle |^2 n_3 (1-n_4)}
{\epsilon_1 + \epsilon_3 - \epsilon_2 - \epsilon_4},
\label{order2qp}
\end{eqnarray}
where the quantity $\bar V = (1-P_{12})V$ denotes the antisymmetrized two-body
potential and $n_j=\theta(k_F-|\vec k_j|)$ is the usual zero-temperature 
Fermi distribution. These calculations revealed the importance of second-order 
diagrams in raising the quasiparticle effective mass from $M^* \simeq 0.7 M_N$ 
to $M^* \simeq M_N$ (both lying in the phenomenological range $M^*/M = 0.7-1.0$
\cite{jeukenne76,zuo99})
as well as increasing the symmetry energy from $\beta \simeq (21-25)$ MeV to 
$\beta \simeq (31-33)$ MeV (where only the second-order result lies within the range of empirical values 
$\beta = (30-36)$\, MeV \cite{danielewicz,steiner}). The isotropic spin-isospin Landau parameter $G_0^\prime$ 
increases at second order, with the effect that $g_{NN}^\prime$ changes from
a value of around $0.50-0.57$ to about $0.67-0.77$, the latter being within 
the range $g_{NN}^\prime = 0.6-0.7$ favored by fits to giant Gamow-Teller resonances in heavy nuclei 
\cite{gaarde83,ericson88,suzuki99}. Despite these encouraging results, the
Fermi liquid parameter $F_0$ remained well below $-1$, giving rise to a 
negative compression modulus ${\cal K}$ and a corresponding instability of nuclear matter 
against density fluctuations in the vicinity of saturation density $\rho_0
=0.16$\,fm$^{-3}$. 
Additionally, the anomalous orbital $g$-factor decreased from $\delta g_l 
\simeq 0.3$ to $\delta g_l \simeq 0.1$, which is significantly less than the 
empirical value
$\delta g_l = 0.20 - 0.26$ extracted from giant dipole resonances \cite{nolte}. 
This feature followed almost entirely from the dramatic
increase in the effective mass $M^*$ at second order. An improved microscopic 
description of the quasiparticle interaction may require the consistent
implementation of chiral three-nucleon forces. As a first step in
this program, we compute here the first-order contribution to
the quasiparticle interaction from the N$^2$LO chiral three-body force.

According to eq.(\ref{deltae}) the quasiparticle interaction is obtained 
by functionally differentiating the energy density twice with respect to
the quasiparticle distribution functions. For a general three-nucleon force, the
Hartree-Fock contribution to the energy density is given by
\be
{\cal E}^{(1)}_{3N} = \frac{1}{6}{\rm tr}_{\sigma_1 \tau_1}
{\rm tr}_{\sigma_2 \tau_2}{\rm tr}_{\sigma_3 \tau_3} \int 
\frac{d^3k_1}{(2\pi)^3} \frac{d^3k_2}{(2\pi)^3} \frac{d^3k_3}{(2\pi)^3}
n_{\vec{k}_1} n_{\vec{k}_2} n_{\vec{k}_3} \langle 123 | \bar{V}_{3N} | 123 
\rangle \, ,
\ee
where $\bar{V}_{3N} = V_{3N}(1-P_{12}-P_{23}-P_{13}+P_{12}P_{23}+P_{13}P_{23})$
denotes the fully antisymmetrized three-nucleon interaction and $n_{\vec k_j}
=\theta(k_F-|\vec k_j|)+(2\pi)^3 \delta^3(\vec k_j-\vec p_j)\, 
\delta n_{\vec p_j \sigma_j \tau_j}$. 
Functionally differentiating 
twice with respect to the two quasiparticle distribution functions then leaves 
an effective two-body interaction containing a single (loop) integral over the 
filled Fermi sea of nucleons.

The three-nucleon force employed in the present work is the N$^2$LO 
chiral three-nucleon interaction \cite{epelbaum06}, which consists of three 
components. First, there is a two-pion exchange component
\begin{equation}
V_{3N}^{(2\pi)} = \sum_{i\neq j\neq k} \frac{g_A^2}{8f_\pi^4} 
\frac{\vec{\sigma}_i \cdot \vec{q}_i \, \vec{\sigma}_j \cdot
\vec{q}_j}{(\vec{q_i}^2 + m_\pi^2)(\vec{q_j}^2+m_\pi^2)}
F_{ijk}^{\alpha \beta}\tau_i^\alpha \tau_j^\beta,
\label{3n1}
\end{equation}
where $g_A=1.29$, $f_\pi = 92.4$ MeV, $m_{\pi} = 138$ MeV (average 
pion mass) and $\vec{q}_i$ denotes
difference between the final and initial momenta of nucleon $i$. The quantity 
\begin{equation}
F_{ijk}^{\alpha \beta} = \delta^{\alpha \beta}\left (-4c_1m_\pi^2
 + 2c_3 \vec{q}_i \cdot \vec{q}_j \right ) + 
c_4 \epsilon^{\alpha \beta \gamma} \tau_k^\gamma \vec{\sigma}_k
\cdot \left ( \vec{q}_i \times \vec{q}_j \right )
\label{3n4}
\end{equation}
involves three terms proportional to the low-energy constants $c_1, c_3$ and 
$c_4$, respectively. The summation $\sum_{i\neq j\neq k}$ runs over
the six permutations of three nucleons.
The one-pion exchange component of the three-nucleon interaction is
proportional to the low-energy constant $c_D$:
\begin{equation}
V_{3N}^{(1\pi)} = -\sum_{i\neq j\neq k} \frac{g_A c_D}{8f_\pi^4 \Lambda_\chi} 
\frac{\vec{\sigma}_j \cdot \vec{q}_j}{\vec{q_j}^2+m_\pi^2} \vec{\sigma}_i \cdot
\vec{q}_j \, {\vec \tau}_i \cdot {\vec \tau}_j \, ,
\label{3n2}
\end{equation}
and finally the three-nucleon contact interaction introduces the low-energy constant $c_E$:
\begin{equation}
V_{3N}^{(\rm ct)} = \sum_{i\neq j\neq k} \frac{c_E}{2f_\pi^4 \Lambda_\chi}
{\vec \tau}_i \cdot {\vec \tau}_j\, ,
\label{3n3}
\end{equation}
where $\Lambda_{\chi} = 700$ MeV sets a natural scale. 

The low-energy constants of $V_{3N}^{(2\pi)}$ appear already in the 
two-pion exchange contribution to the nucleon-nucleon interaction and
can therefore be fit to NN scattering phase shifts. The analysis of the
Nijmegen group \cite{rentmeester03} resulted in the values 
$c_1 =-0.76\,$GeV$^{-1}$, $c_3=-4.78\,$GeV$^{-1}$, and $c_4 =3.96\,$GeV$^{-1}$,
while Entem and Machleidt \cite{entem03} obtain $c_1 =-0.81\,$GeV$^{-1}$, 
$c_3=-3.2\,$GeV$^{-1}$, and $c_4 =5.4\,$GeV$^{-1}$. The low-energy constants
$c_D$ and $c_E$ of the mid- and short-range chiral three-nucleon interaction can be
constrained by properties of nuclear few-body systems. In the present 
work we employ two versions of $c_D$ and $c_E$ obtained by fitting
the binding energies of $A=3$ nuclei together with the half-life of $^3$H 
with the result \cite{gazit}:
\be
c_D = -0.20, \hspace{.3in} c_E = -0.205\, ,
\ee
or by fitting the binding energies of $^3$H and $^4$He \cite{bogner05}:
\be
c_D = -2.062, \hspace{.3in} c_E = -0.625.
\ee
In the first set the leading-order three-body force 
(with $c_{1,3,4}$ coefficients of Entem and Machleidt) was used together with the
Idaho N$^3$LO chiral two-nucleon interaction, while in the second 
set the authors employed the Nijmegen low-energy constants $c_{1,3,4}$
together with the low-momentum NN interaction \vlk at the resolution scale
$\Lambda = 2.1$ fm$^{-1}$. Employing these two versions of 
the chiral three-nucleon force (combined with two-nucleon interactions at
different scales) provides a means for assessing theoretical
errors at this order in the perturbative expansion.

\subsection{Diagrammatic calculation}

We begin by considering the isotropic ($L=0$) Fermi liquid parameters of the central 
quasiparticle interaction. They are obtained by angle-averaging the (in-medium) effective 
interaction \cite{holt10}:
\begin{eqnarray} 
{\cal F}_0(k_f) &=& {1\over (4\pi)^2}\!\int\!d\Omega_1
d\Omega_2\,\langle  \vec p_1, \vec p_2 | V_{NN}^{\rm med}| \vec p_1,\vec p_2
\rangle \nonumber  \\ &=&  f_0(k_f) + g_0(k_f) \, \vec \sigma_1\cdot\vec
\sigma_2 + f'_0(k_f)\, \vec \tau_1 \cdot\vec \tau_2 + g'_0(k_f) \, \vec
\sigma_1 \cdot\vec \sigma_2\, \vec \tau_1 \cdot \vec \tau_2 \, , 
\end{eqnarray}
where $| {\vec p}_1 {\vec p}_2 \rangle $ denotes an antisymmetrized 
two-nucleon state. Both quasiparticle momenta ${\vec p}_1$ and
${\vec p}_2$ lie on the Fermi surface so that $|\vec p_1|=|\vec p_2| =k_f$.
\begin{figure}
\begin{center}
\includegraphics[scale=1.0,clip]{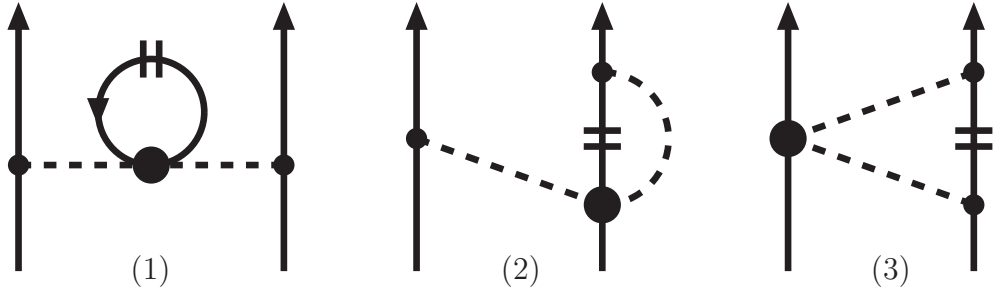}
\end{center}
\vspace{-.5cm}
\caption{Diagrammatic contributions to the quasiparticle interaction in 
symmetric nuclear matter generated from the two-pion exchange
three-nucleon force. The short double-line symbolizes summation over the 
filled Fermi sea of nucleons. Reflected diagrams of (2) and (3) are not shown.}  
\label{mfig1}
\end{figure}
\begin{figure}
\begin{center}
\includegraphics[scale=1.0,clip]{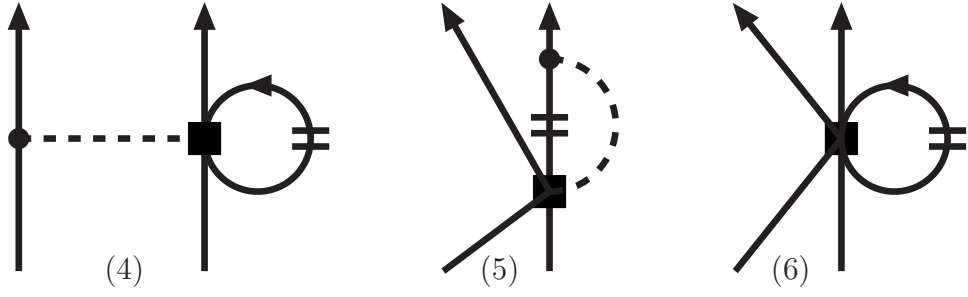}
\end{center}
\vspace{-.5cm}
\caption{Diagrammatic contributions to the quasiparticle interaction in 
symmetric nuclear matter generated by one-pion exchange
and contact three-nucleon forces. Reflected diagrams of (4) and (5) are not 
shown. The last two diagrams (5) and (6) contribute only to the $L=0$ Landau 
parameters.}
\label{mfig2}
\end{figure}
There are six topologically distinct one-loop diagrams, shown in Figs.\ 
\ref{mfig1} and \ref{mfig2}, contributing to the effective interaction 
$V_{NN}^{{\rm med}}$. Using the abbreviations
$\boldsymbol\sigma= \vec \sigma_1 \cdot\vec
\sigma_2$ and $\boldsymbol\tau= \vec \tau_1 \cdot
\vec\tau_2$, we find for the crossed 
term of diagram (1) in Fig.\ \ref{mfig1} the following contribution
\begin{equation}
{\cal F}_0(k_f)^{(\rm med,1)} = (3-\boldsymbol\sigma)(3-\boldsymbol\tau) {g_A^2 m_\pi^3  \over (6\pi)^2 
f_\pi^4} \bigg\{ {(2c_1-c_3)u^3 \over 1+4u^2} -c_3 u^3 +(c_3-c_1) {u \over 2}
\ln(1+4u^2) \bigg\}\,, 
\end{equation}
with $u = k_F/m_\pi$. In this diagram, the two pions carry equal momenta and
therefore the term in eq.(\ref{3n4}) proportional to $c_4$ does not contribute. 
Moreover, the direct term (with zero
momentum transfer) of this pion self-energy correction vanishes trivially.
Analogously, only the crossed term from the one-pion exchange vertex correction 
(diagram (2) in Fig.\ \ref{mfig1}) is nonzero with the analytical result:
\begin{eqnarray}
{\cal F}_0(k_f)^{(\rm med,2)} &=& (3-\boldsymbol\sigma)(3-\boldsymbol\tau) {g_A^2 m_\pi^3  \over
(24\pi)^2 f_\pi^4} \bigg\{{3c_1 \over 8u^5}\Big[4u^2- \ln(1+4u^2)\Big]
\nonumber \\ 
& \times& \Big[8u^4+4u^2-(1+4u^2) \ln(1+4u^2)\Big]\nonumber \\ 
&+& c_3 \bigg[ {2\over u^2} \Big(4u^2- \ln(1+4u^2)\Big)\arctan 2u  
+{48u^4+16u^2 +3 \over 64u^7}  \ln^2(1+4u^2) \nonumber \\ & +& {12u^4-16u^6
-30u^2-9 \over 24 u^5} \ln(1+4u^2) +{20u^3 \over 3} -11 u +{1\over u} + 
{3 \over 4u^3} \bigg] \nonumber \\ & +& c_4 \bigg[ {4\over u^2} 
\Big(\ln(1+4u^2)-4u^2\Big)\arctan 2u +{3+16u^2-48u^4 \over 64u^7} 
\ln^2(1+4u^2) \nonumber \\ & +& {80u^6+12u^4-30u^2 -9 \over 24 u^5} 
\ln(1+4u^2) -{28u^3 \over 3} +13 u +{1\over u} +{3 \over 4u^3} \bigg] 
\bigg\}\, .
\end{eqnarray}
The Pauli-blocked two-pion exchange component (diagram (3) in Fig.\
\ref{mfig1}) has both a nonvanishing direct and crossed term. Their sum takes the
form 
\begin{eqnarray}   
{\cal F}_0(k_f)^{(\rm med,3)} &=& {g_A^2 m_\pi^3 \over (4\pi)^2
f_\pi^4} \bigg\{ 24(c_3-c_1) u - 8 c_3  u^3  +(3c_3-4c_1){3\over 
u}\ln(1+4u^2) \nonumber \\ & +& 6(6 c_1-5c_3) \arctan 2u+(3-\boldsymbol\sigma)(3-\boldsymbol\tau)\, 
{c_4 \over 9}\! \int_0^u\!\!dx\, (Y^2 -X^2)  \nonumber \\ &+& 
(1+\boldsymbol\sigma)(1+\boldsymbol\tau)\!\int_0^u\!\!dx \Big[ 3c_1 \, Z^2 + {c_3 \over 2} (X^2 + 
2 Y^2) \Big] \bigg\} \,, 
\label{pbtpe}
\end{eqnarray} 
where the auxiliary functions $X, Y,$ and $Z$ arising from Fermi sphere 
integrals over a pion propagator read
\begin{eqnarray} 
X(u,x) &=& 2x-{1\over 2u} \ln{1+(u+x)^2 \over 1+(u-x)^2} \,,\\
 Y(u,x)&=&{5x^2-3u^2-3\over 4 x}+{4x^2+3(1+u^2-x^2)^2\over 16 u 
x^2}\ln{1+(u+x)^2 \over 1+(u-x)^2} \,, \\  
Z(u,x)  &=& 1 +{x^2-u^2-1\over 4u x} \ln{1+(u+x)^2 \over  1+(u-x)^2} \, .
\end{eqnarray}
Note that the direct term contributes only to the spin- and 
isospin-independent Landau parameter $F_0$.

There are two diagrammatic contributions from the mid-range one-pion
exchange chiral three-nucleon force, labeled as (4) and (5) in Fig.\ 
\ref{mfig2}. The crossed term from the $1\pi$-exchange vertex 
correction (diagram (4)) leads to the contribution
\begin{equation}
{\cal F}_0(k_f)^{(\rm med,4)} = (3-\boldsymbol\sigma)(3-\boldsymbol\tau) {g_A c_D m_\pi^3  \over
(24\pi)^2 f_\pi^4 \Lambda_\chi} \Big[ 4u^3 -u  \ln(1+4u^2)\Big]\, ,
\end{equation}
and the sum of direct and crossed terms from diagram (5) yields
\begin{equation}
{\cal F}_0(k_f)^{(\rm med,5)} = (3-\boldsymbol\sigma-\boldsymbol\tau-\boldsymbol\sigma\, \boldsymbol\tau) {g_A c_D  m_\pi^3  
\over (4\pi)^2 f_\pi^4 \Lambda_\chi} \bigg\{ {2u^3 \over 3} -u +\arctan 2u
-{1\over 4u} \ln(1+4u^2)   \bigg\} \, .
\end{equation}
Finally, the three-nucleon contact term generates a contribution 
proportional to the nuclear density $\rho = 2k_F^3/3\pi^2$.
The sum of direct and crossed terms reads
\begin{equation}
{\cal F}_0(k_f)^{(\rm med,6)} = (\boldsymbol\sigma+\boldsymbol\tau+\boldsymbol\sigma\, \boldsymbol\tau -3) {c_E \, k_f^3  
\over 4\pi^2 f_\pi^4 \Lambda_\chi}  \, .
\end{equation}
We note that for all three-body contributions the spin-spin
and isospin-isospin components of the quasiparticle 
interaction are equal, $g_0=f_0^\prime$. 

The $p$-wave ($L=1$) Fermi liquid parameters are obtained by weighting the
angular integrals by the first Legendre polynomial $3P_1(\cos \theta) = 3
\hat p_1 \cdot \hat p_2$: 
\begin{eqnarray} {\cal F}_1(k_f) &=& {3\over (4\pi)^2}\!\int\!d\Omega_1
d\Omega_2\,(\hat p_1 \cdot \hat p_2) \langle  \vec p_1, \vec p_2 | 
V_{NN}^{\rm med}|
\vec p_1,\vec p_2\rangle \nonumber  \\ &=&  f_1(k_f) + g_1(k_f) \, \vec 
\sigma_1
\cdot\vec \sigma_2 + f'_1(k_f)\, \vec \tau_1 \cdot\vec \tau_2 + 
g'_1(k_f) \, \vec
\sigma_1 \cdot\vec \sigma_2\, \vec \tau_1 \cdot \vec \tau_2 \, .
\end{eqnarray}
After this weighting, the two
short-range contributions (diagrams (5) and (6) in Fig.\ \ref{mfig2})
to the quasiparticle interaction vanish. We provide the expressions for
the four remaining pieces below. The crossed term from the pion self-energy 
correction takes the form
\begin{eqnarray}
{\cal F}_1(k_f)^{(\rm med,1)} &=& (3-\boldsymbol\sigma)(3-\boldsymbol\tau) {g_A^2 m_\pi^3\over
48\pi^2 f_\pi^4} \bigg\{ {(2c_1-c_3)u \over 1+4u^2}+(6c_1-5c_3)u \nonumber \\
& +& \Big[ 2(c_3-c_1)u +{3c_3-4c_1 \over 2u}\Big] \ln(1+4u^2) \bigg\}\, .
\end{eqnarray}
The crossed term from the one-pion exchange vertex correction reads
\begin{eqnarray}
{\cal F}_1(k_f)^{(\rm med,2)} &=& (3-\boldsymbol\sigma)(3-\boldsymbol\tau) {g_A^2 m_\pi^3  \over
(16\pi)^2 f_\pi^4} \bigg\{{c_1 \over 4u^7}\Big[4u^2- (1+2u^2)\ln(1+4u^2)\Big]
\nonumber \\ 
& \times & \Big[8u^4+4u^2-(1+4u^2) \ln(1+4u^2)\Big] \nonumber \\ &+&
{c_3\over 3u^4}\bigg[ 4 \Big(4u^2- (1+2u^2) \ln(1+4u^2)\Big) \arctan 2u
\nonumber \\ & +& {96u^6+80u^4+22u^2+3 \over 32u^5}\ln^2(1+4u^2) 
\nonumber \\ & +& {56u^6 -32u^8-60u^4-48u^2-9 \over 12u^3} \ln(1+4u^2) \nonumber \\ 
& +&{4u^5 \over 3}(7-4u^2) -14 u^3 +5u +{3 \over 2u} \bigg] \nonumber \\ 
& +& {c_4\over 3u^4}\bigg[8\Big( (1+2u^2) \ln(1+4u^2)-4u^2\Big)\arctan 2u 
\nonumber
\\ &+& {3+22u^2-16u^4 -96u^6\over 32u^5}  \ln^2(1+4u^2) \nonumber \\ 
& +& {160 u^8 -136u^6-60u^4-48u^2 -9 \over 12 u^3} \ln(1+4u^2)  \nonumber \\ 
&-& {4u^5 \over 3}(4u^2+17) +34 u^3 +5u +{3 \over 2u} \bigg] \bigg\}\, .
\end{eqnarray}
The crossed term from Pauli-blocked two-pion exchange is given by
\begin{eqnarray}
{\cal F}_1(k_f)^{(\rm med,3)} &=& {g_A^2 m_\pi^3 \over (4\pi)^2
f_\pi^4}\int_0^u\!\!dx  \bigg\{(3-\boldsymbol\sigma)(3-\boldsymbol\tau)\,{c_4 \over 9} (3X_b^2 + 
2X_c^2-3X_a^2) \nonumber \\ 
&+& (1+\boldsymbol\sigma)(1+\boldsymbol\tau) \Big[3c_1(Z_a^2+2Z_b^2) + {c_3\over 2}
(3X_a^2 + 6X_b^2+4X_c^2) \Big]  \bigg\} \,, \end{eqnarray}
with auxiliary functions
\begin{eqnarray} Z_a(u,x) &=& {x\over u}+{u^2-x^2-1\over 4u^2} 
\ln{1+(u+x)^2
\over 1+(u-x)^2} \,, \\ Z_b(u,x)&=&{x^2-3u^2-3\over 4 u 
x}+{4(1+u^2+u^4)-(u^2
+x^2-1)^2\over 16 u^2 x^2}\ln{1+(u+x)^2 \over 1+(u-x)^2} \,, \\  
X_a(u,x)  &=&
{1\over u} -{u^2+x^2+1\over 4u^2 x} \ln{1+(u+x)^2 \over  1+(u-x)^2} \,,
\\  X_b(u,x)  &=& {1\over 8u x^2}\Big[2x^2(1+u^2)-3x^4-3(1+u^2)^2\Big]
\nonumber \\ & +& {1+u^2+x^2\over 32u^2 x^3}\Big[3(1+u^2-x^2)^2+4x^2\Big]
\ln{1+(u+x)^2 \over  1+(u-x)^2}\,, \\  X_c(u,x) &=& {1\over 
8ux^2}\Big[3(1+u^2)^2
-8u^2 x^2 -3x^4 \Big]\nonumber \\ & +& {3\over 32u^2 x^3}(x^2-1-u^2)\Big[
(1+u^2+x^2)^2-4u^2x^2\Big] \ln{1+(u+x)^2 \over  1+(u-x)^2}  \, . 
\end{eqnarray}
Finally, the only nonvanishing term from the mid-range three-nucleon force 
is the crossed term from diagram (4) in Fig.\ \ref{mfig2},
which reads
\begin{equation}
{\cal F}_1(k_f)^{(\rm med,4)} = (3-\boldsymbol\sigma)(3-T) {g_A c_D m_\pi^3  \over
384\pi^2 f_\pi^4 \Lambda_\chi} \bigg\{ 4u- \Big(2u+{1\over u}\Big) 
\ln(1+4u^2) \bigg\}\, .
\end{equation}
Again, one observes that for all contributions, $f_1^\prime = g_1$.

This completes the first-order calculation of the $L=0,1$ Landau parameters
arising from the N$^2$LO chiral three-nucleon force. A good check of the
formulas in eqs.(17)--(35) is given by their Taylor expansions in $k_F$. The
leading terms are of the form: $c_E k_F^3, c_D k_F^5, c_1 k_F^5, c_{3,4} k_F^7$.
Note that full consistency with 
the second-order calculation in ref.\ \cite{holt11} would require the
inclusion of the subleading N$^3$LO chiral three-body force (recently derived
in ref.\ \cite{bernard08}). Investigations along these lines are in progress.


\section{Results}
\label{results}

In this section we study the density-dependence of the $L=0,1$ Landau parameters
derived in the previous section from the 
leading-order chiral three-nucleon interaction. The values of the five 
low-energy constants occuring at this order have significant uncertainties, and
we employ two different sets which have been fit
to reproduce properties
of $A=3,4$ nuclei. These contributions are then combined with the
results of ref.\ \cite{holt11} for the second-order quasiparticle 
interaction computed with the corresponding two-nucleon interactions. The 
impact on various nuclear observables is discussed.

\begin{figure}
\begin{center}
\includegraphics[height=14cm,angle=270]{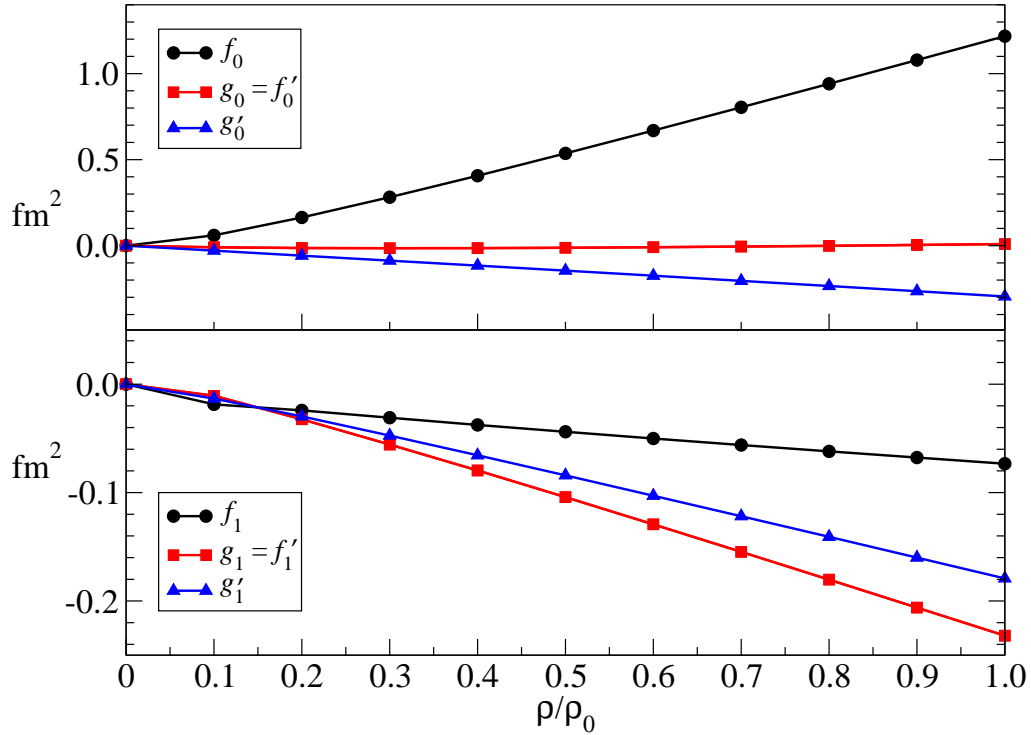}
\end{center}
\vspace{-.5cm}
\caption{Density-dependent $L=0,1$ Fermi liquid parameters obtained from the 
leading-order chiral three-nucleon force with low-energy constants given in
eq.\ (\ref{lecn3lo}).}
\label{DFLPn3lo}
\end{figure}

In Fig.\ \ref{DFLPn3lo} we plot the $L=0,1$ Fermi liquid parameters (in units 
of fm$^2$) as a function of the nuclear density (normalized to that of
saturated nuclear matter $\rho_0 = 0.16\,$fm$^{-3}$) employing the set of 
low-energy constants
\be
c_1 =-0.81\, {\rm GeV}^{-1}, \hspace{.1in} c_3=-3.2\,{\rm GeV}^{-1}, 
\hspace{.1in} c_4 =5.4 \, {\rm GeV}^{-1}, \hspace{.1in} c_D = 
-0.20, \hspace{.1in} c_E = -0.205,
\label{lecn3lo}
\ee
which have been used with the Idaho N$^3$LO chiral two-nucleon interaction
to reproduce the binding energies of $A=3$ nuclei and the triton half-life.
We observe that for densities greater than $\rho \simeq \rho_0/4$, both the 
$L=0$ and $L=1$ Fermi liquid parameters depend approximately linearly on the 
density, which is not immediately obvious from the analytical expressions given in the
previous section. The largest effect on the quasiparticle interaction is a 
dramatic increase in the isotropic spin- and isospin-independent component, 
$f_0$, which at nuclear matter saturation density is about four times larger
than any of the other contributions to the $L=0$ parameters. In fact, both 
$g_0$ and $f_0^\prime$ remain negligibly small for all densities considered
here, while the parameter $g_0^\prime$ decreases monotonically with the density
and reaches the value $-0.3$\,fm$^2$ at nuclear matter saturation density $\rho_0$.
The $L=1$ Fermi liquid parameters all decrease with increasing density. 
The relatively small change in $f_1$ provides only a small downward correction to
the quasiparticle effective mass $M^*$. This observation, combined with the small
change in $f_0^\prime$ discussed above, indicate that three nucleon forces
have a relatively small effect on the nuclear symmetry energy $\beta$. 
The significant reduction in the parameter $f_1^\prime$ decreases
the nucleon anomalous orbital $g$-factor $\delta g_l$ considerably 
below its empirical value.

In Table \ref{n3lotab} we show the various contributions to the $L=0,1$ Fermi 
liquid parameters arising from the six topologically-distinct diagrams of 
Figs.\ \ref{mfig1} and \ref{mfig2}. These are labeled as $V_{NN}^{{\rm med},\, 
i}$ for $i=1,\dots ,6$. Contributions from the long-range two-pion exchange component
of the three-body force are significantly larger than those from the mid- and
short-range three-body forces. However, since both $V_{NN}^{{\rm med},\, 1}$ 
and $V_{NN}^{{\rm med},\, 2}$ are medium modifications to one-pion exchange but
enter with opposite sign, to a large extent they cancel in all Fermi liquid 
parameters. This leaves the Pauli-blocked two-pion exchange diagram 
$V_{NN}^{{\rm med},\, 3}$ as the dominant contribution, except in the Fermi
liquid parameters $g_0=f_0^\prime$ and $f_1$ where its effects are surprisingly 
small.
\setlength{\tabcolsep}{.07in}
\begin{table}
\begin{tabular}{|c||c|c|c|c|c|c||c|} \hline
\multicolumn{8}{|c|}{$k_F=1.33$\, fm$^{-1}$} \\ \hline
 & $V_{NN}^{{\rm med},\, 1}$ & $V_{NN}^{{\rm med},\, 2}$ & $V_{NN}^{{\rm med},\, 3}$ & 
$V_{NN}^{{\rm med},\, 4}$ & $V_{NN}^{{\rm med},\, 5}$ & $V_{NN}^{{\rm med},\, 6}$ & Total \\ 
\hline  \hline
$f_0$ [fm$^2$]& 0.976 & $-$1.196 & 1.277 & $-$0.018 & $-$0.036 & 0.215 & 1.218 \\ \hline
$g_0 = f_0^\prime$ [fm$^2$] & $-$0.325 & 0.399 & $-$0.011 & 0.006 & 0.012 & $-$0.072 & 0.009 \\ \hline
$g_0^\prime$ [fm$^2$] & 0.108 & $-$0.133 & $-$0.209 & $-$0.002 & 0.012 & $-$0.072 & $-$0.295 \\ \hline \hline
$f_1$ [fm$^2$] & $-$0.359 & 0.216 & 0.064 & 0.005 & 0 & 0 & $-$0.073  \\ \hline
$g_1=f_1^\prime$ [fm$^2$] & 0.120 & $-$0.072 & $-$0.278 & $-$0.002 & 0 & 0 & $-$0.232  \\ \hline
$g_1^\prime$ [fm$^2$] & $-$0.040 & 0.024 & $-$0.164 & 0.001 & 0 & 0 & $-$0.179 \\ \hline \hline
\end{tabular}
\caption{Individual contributions to the $L=0,1$ Fermi liquid parameters 
arising from the six diagrams in Figs.\ \ref{mfig1} and \ref{mfig2}. The 
contributions are evaluated at nuclear matter saturation density $\rho_0 =0.16$
fm$^{-3}$ corresponding to a Fermi momentum of $k_F=1.33$ fm$^{-1}$. The values
of the low-energy constants of the leading-order chiral three-nucleon force are
those given in eq.\ (\ref{lecn3lo}).}
\label{n3lotab}
\end{table}
It is instructive to compare the final results to what would be obtained in the
chiral limit ($m_{\pi} \rightarrow 0$):
\begin{eqnarray}
(f_0)^\chi = 1.287\, {\rm fm}^2, \hspace{.2in} (g_0)^\chi = (f_0^\prime)^\chi = 0.136\, {\rm fm}^2, 
\hspace{.2in} (g_0^\prime)^\chi =-0.424\, {\rm fm}^2 \nonumber \\
(f_1)^\chi = -0.073\, {\rm fm}^2, \hspace{.2in} (g_1)^\chi = (f_1^\prime)^\chi = -0.268\, {\rm fm}^2,
\hspace{.2in} (g_1^\prime)^\chi =-0.203\, {\rm fm}^2,
\end{eqnarray}
which are not very different from those at the physical pion mass.

\begin{figure}
\begin{center}
\includegraphics[height=14cm,angle=270]{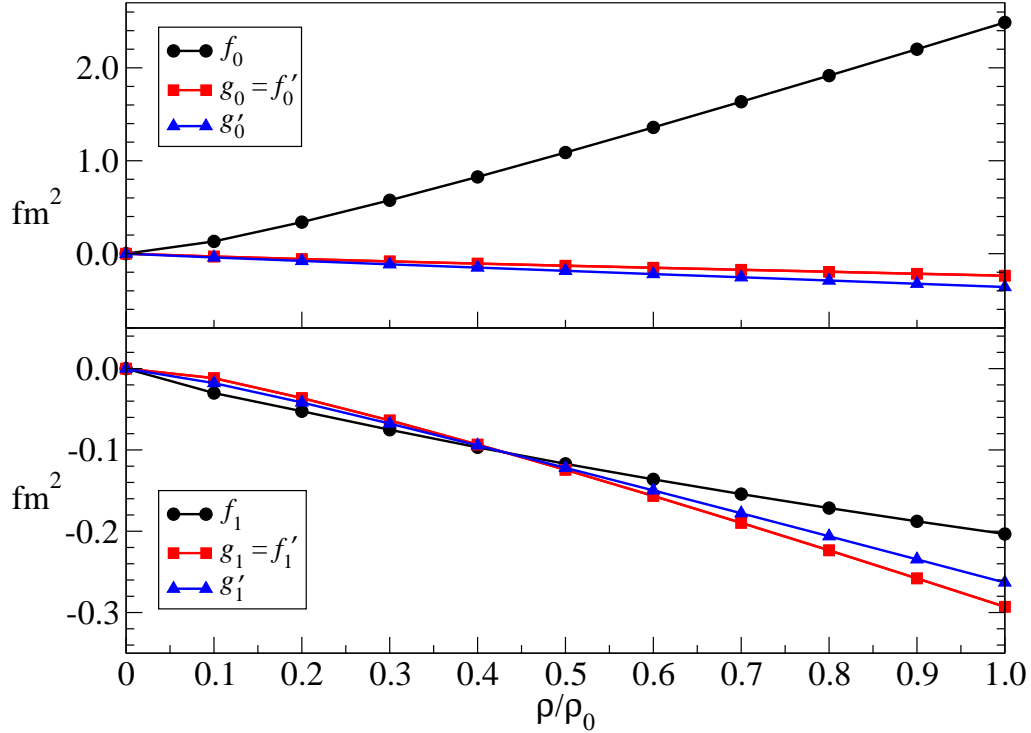}
\end{center}
\vspace{-.5cm}
\caption{Density-dependent $L=0,1$ Fermi liquid parameters obtained from the 
leading-order chiral three-nucleon force with low-energy constants given in
eq.\ (\ref{lecvlk}).}
\label{DFLPvlk}
\end{figure}

To assess the scale dependence of the Fermi liquid parameters resulting from 
the choice of momentum cutoff in the nuclear interaction, we consider next the
set of low-energy constants employed in ref.\ \cite{bogner05} to fit 
$A=3,4$ binding energies using in addition a low-momentum NN interaction 
$V_{\rm low-k}$ defined at the resolution scale $\Lambda = 
2.1$\,fm$^{-1}$:
\be
c_1 =-0.76\, {\rm GeV}^{-1}, \hspace{.1in} c_3=-4.78\,{\rm GeV}^{-1}, 
\hspace{.1in} c_4 =3.96 \, {\rm GeV}^{-1}, \hspace{.1in} c_D = 
-2.06, \hspace{.1in} c_E = -0.625.
\label{lecvlk}
\ee 
In Fig.\ \ref{DFLPvlk} we plot the density-dependent Fermi liquid parameters
resulting from the first-order perturbative contribution from the chiral
three-nucleon force. Qualitatively, many of the trends observed with the 
previous set of low-energy constants (eq.\ (\ref{lecn3lo})) remain. However,
the value of the isotropic spin-independent and isospin-independent Fermi
liquid parameter $f_0$ is significantly larger. At nuclear matter saturation 
density, its value of $f_0= 2.49$\,fm$^2$ is twice as large as that observed
in Fig.\ \ref{DFLPn3lo}. This repulsion more than compensates for the 
additional attraction that results from employing low-momentum two-nucleon interactions
 \cite{holt11}. The parameters $g_0$ and $f_0^\prime$, 
which were nearly zero across the wide range of densities shown in Fig.\ 
\ref{DFLPn3lo}, now achieve a value of $-0.24$\,fm$^2$
at saturation density. All of the $L=1$ Fermi liquid parameters exhibit a 
similar density dependence, and at saturation density lie within the range
$-0.2$\,fm$^2$ to $-0.3$\,fm$^2$. The Fermi liquid parameter $f_1$,
determining the quasiparticle effective mass $M^*$, is twice as large as that 
obtained using the low-energy constants given in eq.\ (\ref{lecn3lo}) and
thus gives rise to a more visible reduction of $M^*$.

We display in Table \ref{vlowktab} the decomposition of the Fermi liquid 
parameters into contributions from the six unique
topologies show in Figs.\ \ref{mfig1} and \ref{mfig2}. Although the 
mid- and short-range components of the 
three-nucleon force are now larger in magnitude, taken together they largely
cancel (in the $L=0$ FLPs) and again give a relatively small contribution to the total sum. 
The largest differences between the values given in Tables \ref{n3lotab}
and \ref{vlowktab} arise from the parameter $c_3$. Employing the low-energy
constants in eq.\ (\ref{lecn3lo}) except with $c_3=-4.78$ would yield the
Fermi liquid parameters
\begin{eqnarray}
f_0 = 2.237\,{\rm fm}^2, \hspace{.2in} g_0 = f_0^\prime = -0.150\,{\rm fm}^2, 
\hspace{.2in} g_0^\prime =-0.341\,{\rm fm}^2 \nonumber \\
f_1 = -0.210\,{\rm fm}^2, \hspace{.2in} g_1 = f_1^\prime = -0.293\,{\rm fm}^2,
\hspace{.2in} g_1^\prime =-0.265\,{\rm fm}^2,
\end{eqnarray}
which are very close to those in Table \ref{vlowktab}.

\setlength{\tabcolsep}{.07in}
\begin{table}
\begin{tabular}{|c||c|c|c|c|c|c||c|} \hline
\multicolumn{8}{|c|}{$k_F=1.33$\, fm$^{-1}$} \\ \hline
 & $V_{NN}^{{\rm med},\, 1}$ & $V_{NN}^{{\rm med},\, 2}$ & $V_{NN}^{{\rm med},\, 3}$ & 
$V_{NN}^{{\rm med},\, 4}$ & $V_{NN}^{{\rm med},\, 5}$ & $V_{NN}^{{\rm med},\, 6}$ & Total \\ 
\hline  \hline
$f_0$ [fm$^2$]& 1.412 & $-$1.171 & 2.149 & $-$0.189 & $-$0.369 & 0.655 & 2.488 \\ \hline
$g_0 = f_0^\prime$ [fm$^2$] & $-$0.471 & 0.390 & $-$0.124 & 0.063 & 0.123 & $-$0.218 & $-$0.237 \\ \hline
$g_0^\prime$ [fm$^2$] & 0.157 & $-$0.130 & $-$0.270 & $-$0.021 & 0.123 & $-$0.218 & $-$0.359 \\ \hline \hline
$f_1$ [fm$^2$] & $-$0.570 & 0.394 & $-$0.082 & 0.054 & 0 & 0 & $-$0.203  \\ \hline
$g_1=f_1^\prime$ [fm$^2$] & 0.190 & $-$0.131 & $-$0.333 & $-$0.018 & 0 & 0 & $-$0.293  \\ \hline
$g_1^\prime$ [fm$^2$] & $-$0.063 & 0.044 & $-$0.250 & 0.006 & 0 & 0 & $-$0.263 \\ \hline \hline
\end{tabular}
\caption{Individual contributions to the $L=0,1$ Fermi liquid parameters 
arising from the six diagrams in Figs.\ \ref{mfig1} and \ref{mfig2}. The 
contributions are evaluated at nuclear matter saturation density $\rho_0 =0.16$
fm$^{-3}$ corresponding to a Fermi momentum of $k_F=1.33$ fm$^{-1}$. The values
of the low-energy constants of the leading-order chiral three-nucleon force are
those given in eq.\ (\ref{lecvlk}).}
\label{vlowktab}
\end{table}

\begin{figure}
\begin{center}
\includegraphics[height=14cm,angle=270]{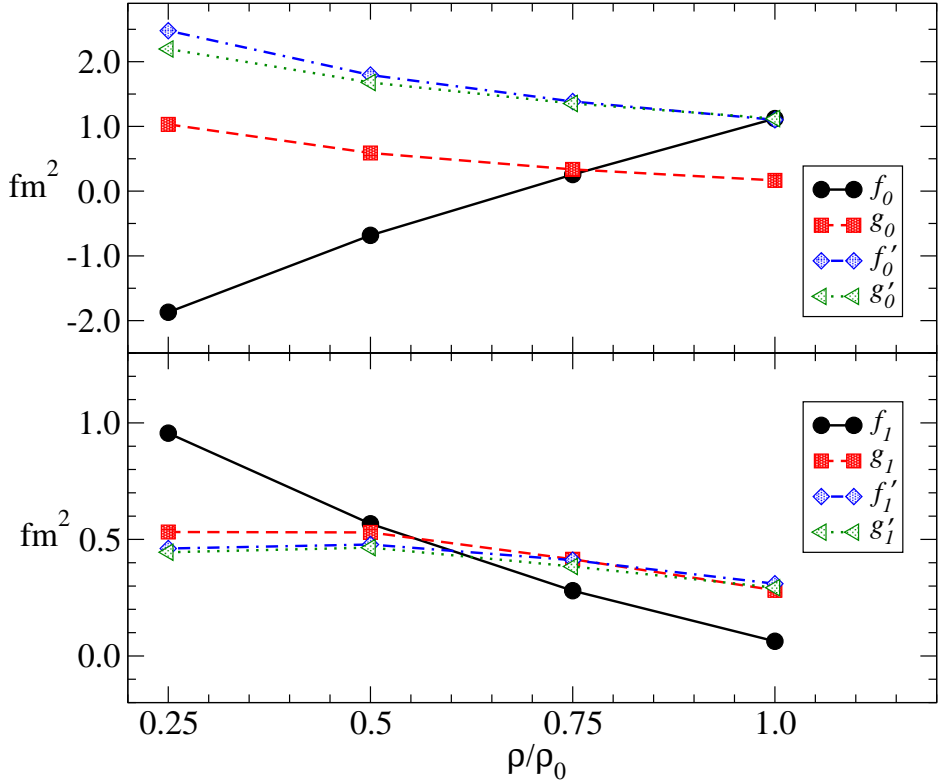}
\end{center}
\vspace{-.5cm}
\caption{Density-dependent $L=0,1$ Fermi liquid parameters. Contributions from 
two-nucleon and three-nucleon forces at the resolution scale $\Lambda =2.1$\,fm$^{-1}$
are added together.}
\label{DDFLP}
\end{figure}

Finally, we combine these results with previous calculations
\cite{holt11} in which the second-order contribution to the quasiparticle
interaction was computed to second order employing chiral and low-momentum NN
interactions. The density dependence of the dimensionful Fermi liquid 
parameters obtained from the low-momentum NN interaction \vlk at a cutoff scale
$\Lambda = 2.1$\,fm$^{-1}$ together with the 
associated three-nucleon force is shown in Fig.\ \ref{DDFLP} (qualitatively similar 
results hold also for the unevolved NN interaction considered in this work). 
Both the $L=0,1$ components of the spin- and
isospin-independent quasiparticle interaction vary strongly with the density, 
and the inclusion of three-nucleon forces significantly enhances the trend 
observed in the second-order calculation including two-body forces
only \cite{holt11} (note that in contrast to the current calculation, the results shown in 
Fig.\ 3 of ref.\ \cite{holt11} did not yet include self-consistent 
single-particle energies in the denominators of the second-order terms). 
The parameters associated with the spin- and isospin-dependent components of 
the quasiparticle interaction, on the other hand, are much more stable to 
variations in the density (in qualitative agreement with observations made in 
ref.\ \cite{kaiser06}). In fact, for these components the three $L=0$ terms
have a remarkably similar density dependence, and the three $L=1$ terms exhibit
this same property.

We now discuss the implications for physical observables. We display in 
Table \ref{finaltable} the dimensionless Fermi liquid parameters (obtained 
by multiplying the dimensionful parameters by the density of states at the 
Fermi surface $N_0=2M^*k_F/\pi^2$) and the related quasiparticle and 
bulk nuclear matter properties resulting from calculations including 
two-nucleon interactions alone as well 
as including the leading-order chiral three-nucleon force. The former are 
denoted with superscript $2N$ while the latter are denoted with $3N$.
The most important feature is a qualitative change of the isotropic 
spin- and isospin-independent Fermi liquid parameter $F_0$, which 
was large and negative without three-body forces but which now attains a mean
value that is positive and relatively small. In fact, the compression modulus
${\cal K}$ computed with the low-momentum interaction is now quite large (${\cal K} \simeq 530$\,MeV)
compared to values ${\cal K}=(200-300)$\,MeV extracted from studies of giant
monopole resonances in heavy nuclei \cite{blaizot,youngblood,ring}.
Given that the use of low-momentum two- and three-body 
interactions yields a quite good description of nuclear matter saturation 
already in (approximate) second order perturbative calculations \cite{bogner05,hebeler11}, it is
likely that the large value of $F_0$ obtained with \vlk in the present work 
results from the absence of second-order contributions from three-body forces.
Aside from the large uncertainty in the value of $F_0$, nearly 
all the other Fermi liquid parameters now 
exhibit much less scale dependence. This feature has also been observed
in previous calculations \cite{bogner05,hebeler11}, where it was found that 
including three-nucleon forces and higher-order corrections in perturbation 
theory dramatically reduced the scale dependence of the nuclear matter
equation of state. The effective mass $M^*$ and the nuclear symmetry energy
$\beta$ receive relatively small corrections from three-body forces at this
order and to within errors remain in agreement with their empirical values.
The observable that is not well described in this calculation is the 
anomalous orbital $g$-factor $\delta g_l$ which remains considerably below the 
value extracted from giant dipole resonances of nuclei. This 
quantity, however, is particularly sensitive to the value of $F_1$ and the 
quasiparticle effective mass (see eq.(\ref{obs})), and therefore even a moderate 
decrease of the nucleon effective mass $M^*$ due to higher-order perturbative
corrections could remedy the situation. We summarize below the results for 
selected nuclear observables, including also theoretical uncertainties estimated 
by choosing different resolution scales and low-energy constants:
\setlength{\tabcolsep}{.07in}
\begin{table}
\begin{tabular}{|c||c|c|c|c||c|c|c|c||c|c|c|c|} \hline
 & $F_0$ & $G_0$ & $F_0^\prime$ & $G_0^\prime$ & $F_1$ & $G_1$ &
$F^\prime_1$ & $G^\prime_1$ & $M^*/M_N$ & ${\cal K}$ [MeV] & $\beta$ [MeV] & $\delta g_l$ \\ 
\hline  \hline
$V_{\rm N3LO}^{2N}$ & $-$1.64 & 0.35 & 1.39 & 1.59 & $-$0.13 & 0.50 & 0.58 & 0.47 
               & 0.96 & $-$150 & 31 & 0.12     \\ \hline
$V_{\rm low-k}^{2N}$ & $-$1.98 & 0.58 & 1.94 & 2.14 & 0.38 & 0.83 & 0.87 & 0.80 
               & 1.13 & $-$190 & 32 & 0.07 \\ \hline \hline
$V_{\rm N3LO}^{3N}$ & $-$0.15 & 0.35 & 1.36 & 1.19 & $-$0.22 & 0.21 & 0.28 & 0.24 
               & 0.93 & 200 & 31 & 0.09     \\ \hline
$V_{\rm low-k}^{3N}$ & 1.48 & 0.22 & 1.45 & 1.48 & 0.08 & 0.37 & 0.41 & 0.39 
               & 1.03 & 530 & 29 & 0.05 \\ \hline
\end{tabular}
\caption{Dimensionless Fermi liquid parameters obtained by summing the 
second-order contribution from the Idaho N$^3$LO potential and low-momentum 
NN interactions $V_{\rm low-k}(\Lambda=2.1\, {\rm fm}^{-1})$ at $k_F=1.33$ fm$^{-1}$. 
Hartree-Fock self-energy insertions, as
parameterized in ref.\ \cite{holt11}, are included in the second-order diagrams.}
\label{finaltable}
\end{table}
\begin{eqnarray}
{\rm Effective \, \, mass:} \hspace{.2in} \frac{M^*}{M_N} =& 0.98 \pm 0.05, &[0.7-1.0]
\nonumber \\
{\rm Anomalous \, \, orbital \, \,} g\mbox{-factor:} \hspace{.2in} \delta 
g_l =& 0.07 \pm 0.02, & [0.20-0.26] \nonumber \\
{\rm Compression \, \, modulus:} \hspace{.2in} {\cal K} =\,\, & (370 \pm 160) \, {\rm MeV},\hspace{.2in}
& [200-300]\, {\rm MeV} \nonumber \\
{\rm Isospin\, \,  asymmetry \, \, energy:} \hspace{.2in} \beta =& 
(30 \pm 1) \, {\rm MeV}, & [30-36]\, {\rm MeV} \nonumber \\
{\rm Spin\mbox{-}isospin\, \, response:} \hspace{.2in} g_{NN}^\prime =& 0.55 \pm 0.03\,
&[0.6-0.7],
\end{eqnarray}
where the quantities in the last column are estimates of the empirical values (see Section
\ref{qp3n}A).
A final interesting feature of the three-nucleon
force is that the contributions to the $L=1$ Fermi liquid parameters are all
negative, with the result that the quasiparticle interaction at the Fermi surface
appears to be short ranged.

\section{Conclusion}
In this work we have performed the first systematic inclusion of the 
N$^2$LO chiral three-nucleon force in calculations of the quasiparticle
interaction in nuclear matter. Analytical expressions for the $L=0,1$ Fermi
liquid parameters were derived, and numerical results 
have been extracted for two different sets of low-energy constants 
($c_E$, $c_D$ and $c_{1,3,4}$) corresponding to two choices of resolution scale
$\Lambda = (500, 414)$\,MeV for the two- and three-nucleon interactions. In comparison to 
calculations employing two-nucleon forces only, we find a significantly smaller 
dependence of the Fermi liquid parameters on the choice of this cutoff scale.
At this order in the perturbative calculation, the bulk properties of nuclear
matter and the dynamical features of quasiparticles are in general well 
described. In particular, the compression modulus ${\cal K}$ of nuclear matter attains a
reasonable value only with the inclusion of three-nucleon correlations, which
introduce large additional repulsion in the effective interaction. Although 
an analogous effect has already been observed in calculations of the nuclear 
matter equation of state, the fact that the three-body force gives a 
disproportionately large contribution to $F_0$, compared to all other Fermi 
liquid parameters, comes as a surprise. The remaining contributions to the 
$L=0,1$ Fermi liquid parameters from the chiral three-nucleon force are 
(essentially) all negative. This results in diminished $L=1$ Landau parameters
and an effective interaction of apparent short range. A remaining unresolved problem 
concerns the description of the anomalous orbital $g$-factor $\delta g_l$ (which arises
from interactions with the nuclear medium). In the present work we find corrections to 
free space orbital $g$-factor that are much too small, a feature which mainly 
results from a large nucleon effective mass ($M^* \simeq M_N$). The role 
played by three-nucleon force contributions beyond the Hartree-Fock
approximation and the effect of the subleading (N$^3$LO) chiral three-body 
force \cite{bernard08} remain topics of interest for future calculations. A
possible strategy is to represent three-nucleon force contributions in terms 
of density-dependent two-body forces \cite{holt10} and to employ these in 
second-order calculations. We have verified that treating three-nucleon force
contributions in this approximation gives good agreement with their
exact evaluation at first order.

\vskip.4in

We thank A.\ Schwenk for helpful discussions.

\end{document}